\documentstyle[aps,manuscript]{revtex}
\begin{document}
\title{Field-driven {\em topological} glass transition \\ in a model
flux line lattice}
\author{Seungoh Ryu}
\address{Dept.~of Physics, Ohio State University, Columbus, OH 43021}
\author {A. Kapitulnik \\ S. Doniach}
\address{Dept.~of Applied Physics, Stanford University, Stanford, CA 94305}
\date{\today}
\maketitle
\begin{abstract} We show that the flux line lattice in a model layered
high temperature superconductor becomes unstable above a critical
magnetic field with respect to a
plastic deformation via penetration of pairs of point-like disclination
defects. The instability is characterized by the competition between the
elastic and the pinning energies and is essentially assisted by softening
of the lattice induced by a dimensional crossover of the fluctuations as
field increases.  We confirm through a computer simulation that this
indeed may lead to a phase transition from crystalline order at low fields
to a topologically disordered phase at higher fields.
We propose that this mechanism provides a model of the low temperature
field--driven disordering transition
observed in neutron diffraction experiments on ${\rm Bi_2Sr_2CaCu_2O_8\, }$ single crystals.  
\end{abstract} 

The mixed state in the high temperature superconductors(HTSC) has been studied
from a wide variety of perspectives over the past few years\cite{blatter95}.
Among the interesting issues under debates are 
the question of how vortices freeze\cite{murray90,chudnovsky89} in the so-called vortex glass phase\cite{fisher89} as well as the nature of the melted phase\cite{nelson88,marchetti90}.
In particular, one expects an intricate interplay of the underlying layered nature of the host material and the random point pins in shaping the nature of the frozen phase.

A recent decoration performed on both sides of the
${\rm Bi_2Sr_2CaCu_2O_8\, }$ single crystal suggests that vortex lines maintain their line
integrity at least at very low field\cite{yao95}. 
On the other hand, the rapid disappearence of Bragg peak intensity in the recent small angle
neutron diffraction(SANS)\cite{cubitt93} has been interpreted by the authors in terms of 
decomposition of lines into pancake vortices\cite{clem91}.
A Lindemann-like criterion is usually invoked in discussing such a ``decoupling theory''\cite{schilling93}, but it fails to explain how the transition is brought about, 
if there is one. 

Traditionally, the role of a pinning potential has been treated as 
a perturbation within the harmonic elastic framework. 
In the weak pinning limit, the perturbed
lattice tries to optimize its free energy by forming  
elastic domains of
correlated region with minimum elastic energy while it is fragmented in larger
length scales to take advantage of the random potential
energy\cite{imry75,larkin79}. 
Several attempts have been made recently to extend the simple
dimensional argument\cite{bouchaud92_3,nattermann90,giamarchi94} for length
scales larger than the elastic volume, but without considerations of
topological defects. 
Their effect is expected to dominate for strong enough disorder, 
resulting in a dislocation dominated `` 
glass phase"\cite{huse94,ryu93a,giam95}.

In an earlier paper\cite{ryu94_1}, we showed that the dramatic features of the SANS results
can not be accounted for by a dimensional crossover alone of a ``clean"
lattice. We further suggested that the flux lattice in the highly anisotropic
${\rm Bi_2Sr_2CaCu_2O_8\, }$ may suffer an instability against penetration of topological defects and
therefore may deteriorate rapidly across a characteristic flux line
density. More recently, such a transition has been claimed to be observed also
in a ${\rm YBa_2Cu_3O_{7-\delta}\, }$ based superlattice\cite{obara95}.
In this letter, we report computer simulation studies of this transition using 
simulated annealing on a model flux line system.
We start by noting that
when the density of flux line in a layered HTSC is varied, the effective
anisotropy of the lattice and the disorder strength change.  As a
result, the elastic domains, described by an in-plane correlation length $R_d$
and an out of plane correlation length $L_d$ may shrink.
In particular, when $L_d$ reaches the interlayer spacing, a new low temperature
 glass state that is dominated by disclinations appears. 
We propose that the transition to this phase is first order and is 
characterized by an explosive invasion of point-like disclination pairs across a horizontal line in the B-T phase diagram.

Employing a vortex representation of the Lawrence-Doniach model
\cite{lawrence70}, we consider a
stack of coupled two dimesional vortex lattices with pancake vortex coordinates
$\{r_{i,z}\}$ with {\em i} labelling the individual flux lines. 
An approximate  pair-wise interaction under periodic boundary conditions has been
derived to make large scale numerical calculations possible\cite{ryu92}.
With a choice of in-plane penetration depth 
$\lambda(0) = 1800\AA, d = 15\AA, \kappa = 100$ and anisotropy $\gamma = \sqrt {M_{c} /  M_{qb}} = 
55,$ we obtain, for a clean system, a melting line
$T_m(B)$ which is in reasonable agreement with the experimental
result\cite{ryu94_5}. The random potential is modelled by potential wells of
uniform depth
$U_p,$ of radius given by the lesser of $2\xi_{ab}$ and the grid size scattered at {\em random positions} of each layer
with an areal density of
$n_p = 1 / a_p^2 \equiv B_p /\phi_0.$ 
We choose $U_p = 5\,{d \phi_0^2 \over 8 \pi^2 \lambda^2(T)}, B_p = 150 G$ in 16(Set \#1) and 32(Set \#3) layers which
give the freezing temperature $ \sim 40 ^\circ K$ for $1 kG < B < 20 kG$
under a Lindemann-like criterion in reasonable agreement with the experimentally determined irreversibility line
$T_{irr}(B).$ We also have results for a different pin density 
$B_p = 2.4 kG$ in 16 layers.(Set \#2) 
The potential used in our simulation may be viewed as a coarse grained
potential fluctuations with an unspecified microscopic origin.
The line density($100 G
\sim 2 kG$) was varied by changing sizes of grid cells and pin densities for a
fixed number of vortices(64 lines in $L=16$ or $32$ layers)\cite{ryu92}.
To obtain low temperature properties, 
we use a {\em simulated annealing} procedure starting
with a perfect  triangular lattice at $T = 1.3\cdot T_{f}(B)\sim 50^\circ K$ 
and gradually decreasing T with  steps of $dT = 5^\circ K$ over
40,000 Monte Carlo steps. At $T=4^\circ K,$ additional 20,000 steps were
performed to make measurement of physical quantities. This formally resembles
a {\em field-cooling} procedure employed in typical measurements. The number of
different pin configurations was limited to five for each line density for
practical reasons. 
The disclination charge density $n_d(i,z)$ is measured every 50
steps through Delaunay triangulation performed in each
layer to determine the coordination number $Z$ of each
vortex\cite{nelson83}. Then a charge of $q = (Z - 6)$ is
assigned to $n_d(i,z)$.

\underline{Results}
In the inset to Fig.~1, we show the partial Fourier transform $S(q_x,q_y,z=L)$
of the vortex density correlation function $ \int d^2 \rho \, \exp [ i q \cdot
\rho ] < n_v(\rho,z=L) \cdot n_v(0,0) >$ as evaluated in our simulation where
$n_v(\rho, z)$ is the local vortex density.
Ideally, SANS with
$q_z \sim 0$ aims to measure
$\int dz S(q_x,q_y,z)$ and therefore for a macroscopic sample($L\rightarrow
\infty)$, we expect it to be related to the diffraction pattern of
\cite{cubitt93}.
The simulated Bragg peak intensities(dots in Fig.~1), decaying rapidly over a narrow
range of $\delta B / B_{cz} \sim 0.3$, looks strikingly similar to the
appropriately scaled data from Cubitt {\em et al}. The values of
$B_{cz}$ were 1800(Set \#1) and 550(\#2) Gauss while we used 500 Gauss for the
experimental data.
The statistical deviation of the intensity, $\delta I_{G1} / I_{G1}$, over
five disorder configurations is within 20 \%. The nature of the 
apparent peak for Set \#1 
in the figure, therefore, can not be resolved at this time.  
To verify what is driving the rapid drop, we look into the behavior of
topological defects and the fraction of pinned vortices, $f_p$ which is given
by $\left< \sum_{i,z} \Theta(r_{i,z} ) \right> / N$ where $\Theta(r)$ gives 1 if
$r$ is within any of the pinning wells and 0 otherwise.  
In Fig.~2, we observe a jump in $f_p$ within the narrow range of $\delta B /
B_{cz} \sim 0.2$, the same region over which the Bragg intensity vanishes.
There is a corresponding increase in the total number of disclinations,
$\sum_{i,z} \left< |n_d(i,z)| \right>$ accompanied by a rapid drop of 
the in-plane hexatic order parameter, $\Psi_6$\cite{hexdef}.
As shown in the inset to the low field side, a typical configuration has a
finite number of defects, but they in general appear as neutral
disclination pairs or quartets(equivalent to a bound pair of dislocations),
healing each other over a finite length along z-axis.
With these bound defects, the lattice order is disrupted only over a finite 
distance given by the size of these bound defects, 
and manifest itself as a distinct Bragg spots of Fig.~1. 
It is to be contrasted
with the higher field configuration in which defects are threading {\em
from-top-to-bottom} layers of the sample in various sizes and separations.
In this phase, we find that the weak hexatic order is still present 
in each layer and that 
their correlation along the c-axis is sensitively dependent on the overall 
strength of the pins\cite{hexcor}.

Introducing an integer variable $n^*_d(i,z)$ which gives 1 for 
a disclination
and 0 otherwise for lattice site $(i,z)$, we define the probability distribution for length $l$ of continuous defect lines:
\begin{equation}
\label{lzdef} {\cal P}(l) = \Big[ \Big< \sum_{i,z}\sum_q [1 - n^*_d(i,z_0)][1 -
n^*_d(i,z_0+l+1)] \prod_{z=z_0+1}^{z_0+l} n^*_d(i,z) \Big> \Big],
\end{equation}
where $< \ldots >$ and $[ \ldots ]$ mean thermal and defect averages, respectively. 
In a pin-free system, the distribution above $T_m(B)$ 
shows a gradual crossover behavior as $B$ increases, 
reflecting the underlying dimensional crossover
of the fluctuations of the lattice\cite{ryu94_5,ryu96}. In this case of
$T \ll T_f$ with quenched disorder, however, we observe a more dramatic
change which suggests a field-driven phase transition. 
In Fig.~3, we show ${\cal P}(l)$ of continuous
defect length $l$ for fields $B/B_{cz} = 0.2 \sim 2$ averaged over five
different pin configurations(Set \#1). 
Despite large statistical fluctuations at the
large tail, we can clearly see that ${\cal P}(l)$ for high field $\sim
e^{-\alpha l}$ qualitatively differs from the low field distribution for which
${\cal P}(l) = 0 $ for $l> l_{max}(B).$
The exponential dependence for high field can easily be understood in terms of
defects generated {\em independently in each layer} with a probability {\em p.}
A probability for an accidentally aligned line of length $l$ will then be 
${\cal P}(l) \sim (1-p)^2 p^{l/d} \sim \exp [ - {l\over d} | \log p | ] .$
In the low field, the length of defect line is truncated by the energetic
balance between the defect rigidity
and the overall gain from random pins.

This observation gives a clue in understanding 
the role of pins in the defect representation of the free energy.
Clearly, the pinning potential acts as an effective ``temperature'' 
to drive the topological defects into
the lattice by encouraging vortices to make large excursions to seek optimal
pinning configurations.
The occurrence of these topological defects signals the breakdown of the 
elastic theory and the transition to a new defect-dominated
phase. We can estimate the breakdown field $B_{cz}$ as follows:
Let us assume that the accumulated vortex displacements $u_i(x)$ 
reach ${\cal O}(a_B)$ over a
volume
$V_d$ of radius $R_d$ in the $ab$-plane, length $L_d$ along the $c$-axis.
For $B < [U_p /
\epsilon d]^4 B_p^2 / B_J,$ where $B_J \equiv \phi_0/(\gamma d)^2,$ a variational minimization of elastic free energy yields $R_d/a_B < 1,$ suggesting a breakdown of the quasi-crystalline order occurring 
in the low field region of the phase diagram. 
Consequently, we assume a very short ranged in-plane order and fix 
$R_d \sim a_B$, and perform a partial variational calculation for a
tube of variable length $L_d|_{R_d \sim a_B}$ to obtain, 
$L_d(B)/d \sim [
\epsilon^2 d^2 B_J^2 / U_p^2 B_p B]^{1/3}.$ The
breakdown field $B_{cz}$ is obtained from the condition
$L_d (B_{cz}) < d.$ Here we assume that the statistical gain in pinning
energy by bending of such tubes is given by $U_p (n_p V_d/d)^{0.5+\delta}
/ (a_B^2L_d)$ with $\delta = 0,$ a rough approximation. Using the
parameter values we chose, this yields $B_{cz} \sim 2.4 {\rm kG,}$[Set
\#1] in reasonable agreement with the result of the simulation.

The results of the simulations may be summarized in a phase diagram given in 
Fig.~4. At low temperatures, the effective ``noise temperature'' of the
random pinning field drives the system from a quasi-crystalline phase at
low fields to a defect-dominated topologically disordered phase at higher 
fields. This phase transition is facilitated by weak interlayer coupling, thus
explaining the small value of the critical field $B_{cz}$ for highly anisotropic
materials such as ${\rm Bi_2Sr_2CaCu_2O_8\, }$. On the other hand, for less anisotropic materials
such as ${\rm YBa_2Cu_3O_{7-\delta}\, }$, this instability should be pushed to a higher field. Because
the disorder is induced by the pins, the vortices are still highly
localized, hence in a glassy phase. We have not yet used our simulations
to study the kinetics of this phase, however, from our previous studies
\cite{ryu93}, we expect the kinetics of the phase to be sub-ohmic, so that
it should still be a superconductor. 

As the temperature is increased in the topologically disordered phase, 
we expect to reach a ``depinning temperature'' above which the defects 
become liquid-like and the transport would be ohmic. 
The exact shape of the line of phase transitions between the quasi-crystalline
phase and the topological glass phase depend on the details of the competing
energies and may terminate at the melting line $T^{3d}_m(B)$ which was
recently shown to bend toward and terminate at a field of $\sim 500\, $ Gauss
\cite{zeldov95}.  In the vicinity of the transition line $B_{cz},$ the
explosion of topological defects will be expected to dominate the
kinetics and may provide an explanation of the ``fishtail'' peak-effect
in the critical current observed as the B-field is varied in this region
of the phase diagram\cite{yeshurun,kam95}.  The results reported here are
reminiscent of a transition observed independently by Gingras and Huse
from  simulations of a ferromagnetic 3d XY model with a random
field\cite{gingras95}.

The authors are grateful for the support by EPRI and Air Force.
SR was also supported by NSF Grant DMR94-02131, 
Midwest Superconductivity
Consortium through DOE Grant DE-FG02-90ER-45427 
and by Ohio State University Postdoctoral Fellowship.
SR acknowledges very useful discussions with Dr.~Forgan and Dr.~Stroud.

\begin{figure}
\caption [first figure] {The relative integrated intensity of the first order
Bragg peak in the partial structure factor for Set \#1 and \#2.
The intensity was
normalized by that of the {\em clean} sample at the same temperature and field.
The broken line is from the Neutron
scattering data\cite{cubitt93} drawn with an arbitrary scale. The field
was arbitrarily scaled by a value at which the intensity reaches the
flat bottom. The insets show the {\em simulated} diffraction pattern for three
representative points(labelled a,b,c) of Set \#1 in the curve. Intensity for (c) was multiplied by a factor of 30.} 
\label{figone}
\end{figure}

\begin{figure}
\caption [second figure] {Pinned fraction of vortices across the instability
line for Set \#1. 
The broken line is a guide showing the naive trend of $\sim B_{p}/B$ as
expected from the relative ratio of the densities. The abrupt increase in the
pinned fraction is closely accompanied by the proliferation of topological
defects as shown as black$(z>6)$ and gray$(z<6)$ dots in the insets.} 
\label{figtwo}
\end{figure}

\begin{figure}
\caption [third figure] {${\cal P}(l)$ for various values of $B$ averaged over
five different realizations of pins(Set \#3). Number of layers was 32. Filled circles
correspond to the hexatic glass region while open symbols are for higher
field topological glass regime. Broken lines are guides for the eye.} 
\label{figthree}
\end{figure}

\begin{figure}
\caption [fourth figure] {Schematic phase diagram with 
{\em  strong} point pins. The low field elastic lattice is turned into a
topological glass state as $B$ increases over $B_{cz}.$ 
The freezing temperatures($T_f^{3d}$ and $T_f^{2d}$) are characterized by the dimensionality of topological defects as discussed in the text.
We also note that there is a distinct decoupling crossover 
line $T_{dc}$ in the liquid phase across which
the vortex lines disintegrate into pancake vortices through cutting and reconnection\cite{monica94}.} 
\label{figfour}
\end{figure}
\break


\begin{thebibliography}{10}
\bibitem{hexdef}
After Delaunay triangulation, the local hexatic order parameter is 
evaluated from $\Psi_6 = { 1\over N} < \sum_{i,j',z} {1 \over c_{i,z}} 
\exp [ i 6 \theta_{ij}(z) ] > $ where $N$ is the total number of pancakes, 
$c_{i,z}$ is the coordination number of a pancake, $\theta_{ij}(z)$ is the
bond angle between neighbouring pancakes measured with respect to a fixed
direction, $\hat{x}.$
\bibitem{hexcor}
For $B_p = 150 G,$ we have $|\Psi_6|^2 \sim 0.2$ with $
<\Psi_6(z) \Psi_6(z-1)^*> \sim 0.2$, while for $B_p = 2.4 kG,$ 
they were $0.1$ and $0.01$ at the field where the Bragg peak 
intensity vanishes. Both $\Psi_6$ and its interlayer correlation 
monotonically decreases as field further increases. 
We note that the kink observed in the pinned fraction around $B\sim 5B_{cz}$
could possibly signal the occurence of an intermediate phase.
\end{thebibliography}
\end{document}